\documentclass[12pt]{article}

\usepackage[makeroom]{cancel}
\usepackage[normalem]{ulem}
\usepackage{amsmath,amssymb}
\usepackage{slashed}
\usepackage{xcolor} 
\usepackage{graphicx}
\usepackage{url}
\usepackage{cite}
\usepackage{pdflscape}
\usepackage{multirow}
\usepackage[thinlines]{easytable}
\usepackage{verbatim}

\newcommand{\fref}[1]{Fig.~\ref{fig:#1}} 
\newcommand{\eref}[1]{Eq.~\eqref{eq:#1}}

\newcommand{\aref}[1]{Appendix~\ref{app:#1}}
\newcommand{\sref}[1]{Section~\ref{sec:#1}}
\newcommand{\cref}[1]{Chapter~\ref{ch:.#1}}
\newcommand{\tref}[1]{Table~\ref{tab:#1}}

\newcommand{\nn}{\nonumber \\}

\newcommand{\beq}{\begin{equation}} 
\newcommand{\eeq}{\end{equation}} 
\newcommand{\ba}{\begin{array}}  
\newcommand{\ea}{\end{array}} 
\newcommand{\bea}{\begin{eqnarray}}  
\newcommand{\eea}{\end{eqnarray} }  
\newcommand{\be}{\begin{eqnarray}}  
\newcommand{\ee}{\end{eqnarray} }  
\newcommand{\bal}{\begin{align}}
\newcommand{\eal}{\end{align}}   
\newcommand{\bi}{\begin{itemize}}  
\newcommand{\ei}{\end{itemize}}  
\newcommand{\ben}{\begin{enumerate}}  
\newcommand{\een}{\end{enumerate}}  
\newcommand{\bc}{\begin{center}}
\newcommand{\ec}{\end{center}} 
\newcommand{\bt}{\begin{table}}
\newcommand{\et}{\end{table}}  
\newcommand{\btb}{\begin{tabular}}
\newcommand{\etb}{\end{tabular}}

\newcommand{\eps}{\epsilon}
\newcommand{\eL}{\epsilon_L}
\newcommand{\eR}{\epsilon_R}
\newcommand{\eS}{\epsilon_S}
\newcommand{\eSL}{\epsilon_{S_L}}
\newcommand{\eSR}{\epsilon_{S_R}}
\newcommand{\eP}{\epsilon_P}
\newcommand{\eT}{\epsilon_T}

\begin{document}

\begin{titlepage}

\vspace*{-2cm}
\begin{flushright}
CERN-TH-2017-103
\vspace*{2mm}
\end{flushright}

\begin{center}
\vspace*{15mm}

\vspace{1cm}
{\LARGE \bf
Renormalization-group evolution of new physics contributions to (semi)leptonic meson decays
} 
\vspace{1.4cm}

\renewcommand{\thefootnote}{\fnsymbol{footnote}}
{Mart\'{i}n~Gonz\'{a}lez-Alonso$^a$, Jorge~Martin~Camalich$^b$, Kin Mimouni$^c$}
\renewcommand{\thefootnote}{\arabic{footnote}}
\setcounter{footnote}{0}

\vspace*{.5cm}
\centerline{$^a${\it IPN de Lyon/CNRS, Universite Lyon 1, Villeurbanne, France}}
\centerline{$^b$ \it CERN, Theoretical Physics Department, Geneva, Switzerland} 
\centerline{$^c$ \it Institut de Th\'eorie des Ph\'enom\`{e}nes Physiques, EPFL, Lausanne, Switzerland}
\vspace*{.2cm}

\end{center}

\vspace*{10mm}
\begin{abstract}\noindent\normalsize

We study the renormalization group evolution (RGE) of new physics contributions to (semi)leptonic charged-current meson decays, focusing on operators involving a chirality flip at the quark level. We calculate their evolution under electroweak and electromagnetic interactions, including also the three-loop QCD running and provide numerical formulas that allow us to connect the values of the corresponding Wilson coefficients from scales at the TeV to the low-energy scales. The large mixing of the tensor operator into the (pseudo)scalar ones has important phenomenological implications, such as the RGE of new physics bounds obtained from light quark decays or in $b\to c\ell\nu$ transitions. For instance, we study scenarios involving tensor effective operators, which have been proposed in the literature to address the $B$-decay anomalies, most notably those concerning the $R_{D^{(*)}}$ ratios. We conclude that the loop effects are important and should be taken into account in the analysis of these processes, especially if the operators are generated at an energy scale of $\sim 1$ TeV or higher.

\end{abstract}

\end{titlepage}
\newpage 

\renewcommand{\theequation}{\arabic{section}.\arabic{equation}} 

\section{Introduction}

Charged-current transitions represent the dominant mechanism for the weak decays of most hadrons and are the main source of experimental inputs to determine the Cabibbo-Kobayashi-Maskawa (CKM) matrix elements. Precision studies of these decays provide benchmarks to test the flavor structure of the Standard Model (SM) and to look for New Physics (NP). A remarkable example is the analysis of (semi)leptonic nuclear, pion and kaon decays, where the very precise database, and exquisite understanding of the SM corrections, lead to sub-permille level tests of the SM that are sensitive to NP scales in the $1-100$ TeV range~\cite{Antonelli:2009ws,Bhattacharya:2011qm,Gonzalez-Alonso:2016etj}. 

Moreover, several tensions with the SM have appeared in the semileptonic decays of $B$ mesons induced by the $b\to (c,~u) \ell\nu$ transitions that could be pointing to the effects of NP. The most significant one corresponds to the $\tau$-to-$\ell$ rate ratios $R_{D^{(*)}}=\Gamma(B\to D^{(*)}\tau\nu)/\Gamma(B\to D^{(*)}\ell\nu)$  (where $\ell$ is a muon or electron) that exhibit enhancements of a 30\% with respect to the SM expectations with a significance of $\sim 4\sigma$~\cite{Lees:2012xj,Lees:2013uzd,Huschle:2015rga,Sato:2016svk,Abdesselam:2016xqt,Aaij:2015yra} (see~\cite{Ciezarek:2017yzh} for a review). Furthermore, \textit{inclusive} and \textit{exclusive} determinations of the CKM matrix element $|V_{cb}|$ and $|V_{ub}|$ obtained with the light-lepton modes have disagreed with each other with a significance in the range of $\sim 2-3\sigma$~\cite{Amhis:2014hma,Olive:2016xmw,Bernlochner:2017jka,Bigi:2017njr,Grinstein:2017nlq}.         

These results have triggered an intense activity in the theory community, both through model-building efforts and through Effective Field Theory (EFT) studies. However, these analyses usually work at tree level, neglecting Renormalization-Group Evolution (RGE) effects. These necessarily appear when the predictions of the underlying model at the high NP scale are connected with the low scale associated with the experimental measurements. It was recently pointed out the importance of such RGE effects for chirality-conserving operators addressing the $B$ anomalies, since they could generate one-loop contributions to $Z$ and $\tau$ decays, which are very precisely measured~\cite{Feruglio:2016gvd,Feruglio:2017rjo}. Although chirality-flipping operators have also been considered in the literature to explain the various anomalies~\cite{Biancofiore:2013ki,Sakaki:2014sea,Duraisamy:2014sna,Freytsis:2015qca,Becirevic:2016hea,Alonso:2016gym,Colangelo:2016ymy,Li:2016vvp,Bardhan:2016uhr,Bhattacharya:2016zcw,Ivanov:2017mrj,Chen:2017hir}, a study of the RGE effects in such scenarios is absent in the literature. We amend this limitation in this work, calculating the one-loop QED and EW evolution of the complete low-energy EFT Lagrangian, including also the three-loop QCD running. We provide numerical formulas that can be trivially implemented in future analyses, and we illustrate the importance of these effects with several simple applications. We find in particular a large mixing of the tensor operator into the (pseudo)scalar ones, which has important phenomenological implications, including for instance NP scenarios that have been proposed in the literature to address several $B$-decay anomalies. 

This work is organized as follows. In \sref{formalism} we introduce the low- and high-energy EFT Lagrangians, and we calculate the associated anomalous dimension matrices. In \sref{running} we use these results to connect the effective operators at the low-energy scale with the NP scale $\Lambda$. We present some phenomenological implications of our results in \sref{phenomenology}, whereas \sref{conclusions} contains our conclusions.

\section{EFTs and anomalous dimensions}
\label{sec:formalism}
\setcounter{equation}{0} 

Assuming Lorentz invariance and heavy NP physics, $d_j\to u_i \ell \bar{\nu}_\ell$ transitions are described at low energies by the following effective Lagrangian
\bea
{\cal L}_{\rm eff} 
&=&
- \frac{4G^0_F V_{ij}}{\sqrt{2}}\,
\Bigg[
S_{\rm ew} \Big(1 + \eL^{ij\ell}  \Big) \bar{\ell}  \gamma_\mu  P_L   \nu_{\ell} \cdot \bar{u}_i   \gamma^\mu P_L d_j
+  \eR^{ij\ell}  \,   \bar{\ell} \gamma_\mu P_L  \nu_\ell    \ \bar{u}_i \gamma^\mu P_R d_j\nonumber\\
&&+~  \eSL^{ij\ell}\, \bar{\ell}  P_L \nu_{\ell} \cdot \bar{u}_i  P_L d_j
+~  \eSR^{ij\ell}\, \bar{\ell}  P_L \nu_{\ell} \cdot \bar{u}_i  P_R d_j
+ \eT^{ij\ell}    \,   \bar{\ell}   \sigma_{\mu \nu} P_L \nu_{\ell}    \cdot  \bar{u}_i   \sigma^{\mu \nu} P_L d_j
\Bigg]+{\rm h.c.}, 
\label{eq:leff1} 
\eea
where $P_{L/R}=(1\mp \gamma_5)/2$ are the chirality projectors, the indices $i,~j$ ($\ell$) label the quark (lepton) families and $G_F^0$ is the Fermi constant. Notice that the $G_F$ value extracted from muon decay within a SM framework can also be affected by NP effects: $G_F^{exp}=G_F^0 +\delta G_F$. The $S_{\rm ew}$ factor encodes the universal short distance corrections to the semileptonic transitions in the SM~\cite{Sirlin:1981ie,Marciano:1993sh}, and the $\epsilon_\Gamma$ coefficients encode the leading NP contributions to this process. Both $S_{\rm ew}$ and the $\epsilon_\Gamma$ coefficients display renormalization-scale dependence that is to be canceled in the observables by the opposite dependence in the quantum corrections to the matrix elements of the decays.\footnote{The only exception is the $\epsilon_L$ coefficient, which is not scale-dependent because it is defined with respect to the SM contribution.} 
Namely, at one loop, we have
\bea
\label{eq:qedrunning}
\frac{d\,\vec \epsilon(\mu)}{d\log\mu}=   \left( \frac{\alpha_{\rm em}(\mu)}{2\pi}\gamma^T_{\rm em} + \frac{\alpha_s(\mu)}{2\pi}\gamma^T_s \right) \,\vec \epsilon(\mu)~,
\eea
where $\vec\epsilon=(S_{\rm ew},\,\eR,\,\eSR,\,\eSL,\,\eT)$, and $\alpha_{\rm em}$ and $\alpha_s$ are the electromagnetic and strong structure constants. We omit flavor indices in the $\epsilon_\Gamma$ coefficients, as both QED and QCD conserve flavor. The matrices $\gamma_{\rm em}$ and $\gamma_s$ are the corresponding one-loop anomalous dimensions, and the superindex $T$ simply indicates matrix transposition. 

\begin{figure}[h!]
\centerline{\includegraphics[scale=0.80]{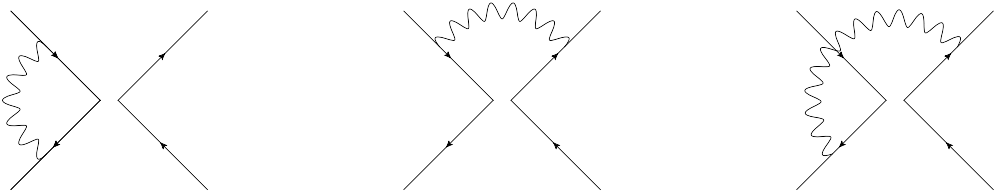}}
\caption{One-loop QED radiative corrections contributing to $d\bar u\to \ell\bar\nu$ effective operators.
\label{fig:diags}}
\end{figure}
 Calculating the diagrams in Fig.~\ref{fig:diags}, we obtain
\bea
\label{eq:gammaqed}
\gamma^T_{\rm em}
=
\left(
\scriptsize{
\begin{array}{ccccc}
y - x_{de} - x_{du} +4 x_{ue} &0&0 & 0 & 0 \\
 0& y - 4x_{de} - x_{du} + x_{ue} &0 & 0 & 0 \\
 0&0 & y - x_{de} - 4x_{du} + x_{ue} & 0 &  0 \\
 0&0&0 &  y - x_{de} - 4x_{du} + x_{ue} &  - 12x_{de} - 12 x_{ue} \\
 0&0& 0 & - \frac{1}{4}x_{de} - \frac{1}{4} x_{ue} &  y - 3x_{de} + 3x_{ue} \\
\end{array}
}\right)\nn\nonumber
\eea
where $y=\sum_{d,u,e}Q_i^2/2$ and $x_{ij}=Q_i Q_j$ (with $Q_f$ denoting the electric charge of the fermion $f$). Numerically,
\bea
\gamma^T_{\rm em}
=
\left(
\begin{array}{ccccc}
 -2&0&0 & 0 & 0 \\
 0&-1&0 & 0 & 0 \\
 0&0 &\frac{2}{3} & 0 & 0 \\
 0&0&0 &  \frac{2}{3} & 4 \\
 0&0&0 &  \frac{1}{12} &  -\frac{20}{9} \\
\end{array}
\right)~,~~~~~\gamma_s =
\left(
\begin{array}{ccccc}
 0&0&0 & 0 & 0 \\
 0&0&0 & 0 & 0 \\
 0&0&-4 & 0 & 0 \\
 0&0& 0 & -4 & 0 \\
 0&0& 0 & 0 & 4/3 \\ 
\end{array}
\right)~.\label{eq:anomalous} 
\eea
The $\gamma_s$ matrix given above is the QCD one-loop result~\cite{Eichten:1989zv}, which is actually known to three loops~(see below). Concerning the electromagnetic anomalous dimension, our result agrees with the well-known SM calculation of the $[\gamma_{\rm em}]_{11}$ element~\cite{Sirlin:1981ie,Marciano:1993sh}, as well as with the partial calculation in Ref.~\cite{Voloshin:1992sn}, where the $[\gamma_{\rm em}]_{54}$ element was calculated. 
%

If the nonstandard particles are not only heavier than the hadronic scale but also much heavier than the electroweak scale, and assuming that the electroweak symmetry breaking is linearly realized, then the low-energy EFT of \eref{leff1} must be matched to the so-called Standard Model EFT (SMEFT)~\cite{Buchmuller:1985jz}
\bea
{\cal L}_{\rm SMEFT} = \sum \frac{w_i}{v^2} {\cal O}_i~,
\eea
where $v=(\sqrt{2}G_F)^{-1/2}\simeq 246$ GeV. 
In this work we will use the so-called Warsaw basis of Ref.~\cite{Grzadkowski:2010es}\footnote{One difference is that for operators with the $SU(2)_L$ singlet contraction of fermionic currents we omit the superscript~${}^{(1)}$.}. \tref{warsaw} lists the relevant operators for this work, which contribute at tree level to the processes $d_j\to u_i\ell\bar{\nu}_\ell$ and $\mu\to e\bar{\nu}_e\nu_\mu$.
\begin{table}[h]
\centering
\caption{Operators in the Warsaw basis~\cite{Grzadkowski:2010es} that are relevant for this work. For further details about the conventions, we refer the reader to the original publication~\cite{Grzadkowski:2010es}.\vspace{0.3cm}
\label{tab:warsaw}}
\begin{tabular}{|c|c|}
\hline
${\cal O}_{\varphi l}^{(3)}$		& $({\varphi^\dag i\,\raisebox{2mm}{${}^\leftrightarrow$} \hspace{-4mm} D_\mu^{\,I}\,\varphi})(\bar l \tau^I \gamma^\mu l)$\\
${\cal O}_{\varphi q}^{(3)}$		& $({\varphi^\dag i\,\raisebox{2mm}{${}^\leftrightarrow$} \hspace{-4mm} D_\mu^{\,I}\,\varphi})(\bar q \tau^I \gamma^\mu q)$\\
${\cal O}_{\varphi u d}$		& $i(\widetilde{\varphi}^\dag D_\mu \varphi)(\bar u \gamma^\mu d)$\\
${\cal O}_{ll}$				& $(\bar l \gamma_\mu l)(\bar l \gamma^\mu l)$\\
\hline
\end{tabular}
\hspace{1cm}
\begin{tabular}{|c|c|}
\hline
${\cal O}_{lq}^{(3)}$			& $(\bar l \gamma_\mu \tau^I l)(\bar q \gamma^\mu \tau^I q)$\\
${\cal O}_{ledq}$			& $(\bar l^j e)(\bar d q^j)$\\
${\cal O}_{\ell equ}$			& $(\bar l^j e) \eps_{jk} (\bar q^k u)$\\
${\cal O}_{\ell equ}^{(3)}$		& $(\bar l^j \sigma_{\mu\nu} e) \eps_{jk} (\bar q^k \sigma^{\mu\nu} u)$\\
\hline
\end{tabular}
\end{table}

Once again, the Wilson coefficients $w_i$ are scale-dependent quantities. Here we focus on the coefficients $\vec w=\left\{w_{ledq},\,w_{\ell equ},\,w^{(3)}_{\ell equ}\right\}$, which are particularly interesting from a phenomenological point of view, as we will explain below. Their running above the EW scale is described by
\bea
\label{eq:ewrunning}
\frac{d\,\vec w(\mu)}{d\log\mu}=   \left( \frac{1}{8\pi^2}\gamma^T_w(\mu) + \frac{\alpha_s(\mu)}{2\pi}{\hat\gamma}^T_s \right) \,\vec w(\mu)~,
\eea
where $\hat\gamma_s$ refers to the $3\times3$ sub-matrix of $\gamma_s$ corresponding to the running of $(\epsilon_{S_R}, \epsilon_{S_L},\epsilon_T)$. We noticed that the result of Ref.~\cite{Alonso:2013hga} for the electroweak anomalous dimension matrix $\gamma_w$ disagrees with a previous calculation~\cite{Campbell:2003ir}. We obtain

\bea
\label{eq:gammaEW}
\gamma^T_{\rm w}=
\left(
\begin{array}{ccc}
 -\frac{4}{3}g_Y^2		&	0					&		0		\\
 0		        	& -\frac{11}{6}g_Y^2	        	        &	15 g_Y^2 + 9 g_L^2  \\
 0				& \frac{5}{16}g_Y^2 + \frac{3}{16}g_L^2         & \frac{1}{9}g_Y^2 - \frac{3}{2}g_L^2 \\
\end{array}
\right)
\approx
\left(
\begin{array}{ccc}
 -0.169 & 0 & 0 \\
 0 & -0.232 & 5.698\\
 0 & 0.119 & -0.619 \\
\end{array}
\right)~,
\eea
which agrees with Ref.~\cite{Alonso:2013hga} neglecting terms suppressed by Yukawa couplings. Let us note that the difference with Ref.~\cite{Campbell:2003ir} does not have any significant impact on RGE calculations, since it only affects the diagonal elements, whose running is fully dominated by QCD. The RHS of \eref{gammaEW} shows for illustration the numerical value of the electroweak anomalous dimension matrix for $\mu=m_Z$. One can see that the mixing is one order of magnitude larger than in the QED case.

\subsection{Higher QCD corrections}

Higher loop corrections to the QCD running of the Wilson coefficients can give a sizable impact when evolving to or from low-energy scales. At three loops we have
\bea
\frac{d\,\vec \epsilon(\mu)}{d\log\mu}=   \left(\frac{\alpha_s(\mu)}{2\pi}\gamma^{(1)}_s + \frac{\alpha_s(\mu)^2}{8\pi^2}\gamma^{(2)}_s + \frac{\alpha_s(\mu)^3}{32\pi^3}\gamma^{(3)}_s \right) \,\vec \epsilon(\mu)~,
\eea
where $\gamma_s^{(n)}$ is the $n$-loop anomalous dimension matrix. The 2- and 3-loop results have the same structure as the 1-loop matrix $\gamma_s^{(1)}$ in \eref{anomalous}, with the following non-zero entries~(see Ref.~\cite{Gracey:2000am} and references therein)
\begin{align}
[\gamma_s^{(2)}]_{33} = [\gamma_s^{(2)}]_{44} = & \frac{2}{9}(-303+10 n_f) \nonumber\\
[\gamma_s^{(3)}]_{33} = [\gamma_s^{(3)}]_{44} = & \frac{1}{81} \left(-101169 + 24 (277+180 \zeta(3)) n_f + 140 n_f^2 \right)  \nonumber\\
[\gamma_s^{(2)}]_{55} = & \frac{2}{27} (543-26 n_f) \nonumber\\
[\gamma_s^{(3)}]_{55} = & \frac{1}{81} \left(52555 - 2784 \zeta(3)-40(131+36\zeta(3))n_f -36 n_f^2 \right) \,.
\end{align}

The running from low- to high-energy renormalization scales might require to cross various heavy-quark thresholds. The so-called threshold corrections that relate the Wilson coefficients in the $n_{f-1}$ and $n_{f}$ effective theories have the following form
\begin{align}
\epsilon_{\Gamma}^{(n_f-1)}(\mu_{th}) = & \epsilon_{\Gamma}^{(n_f)}(\mu_{th}) \left[1 + \sum_n \xi_{\Gamma,n} \left(\frac{\alpha_s^{(n_f)}(\mu_{th})}{\pi} \right)^n \right]~~~~(\Gamma=S,P,T)~. 
\end{align}
The (pseudo)scalar Wilson coefficients have the same QCD running as the quark masses~\cite{Chetyrkin:1997un}
\bea
\xi_{S/P,1}=0 \,, \quad \xi_{S/P,2} = \frac{89}{432} \,.
\label{eq:thrsld_scalar}
\eea

The matching coefficients for the tensor Wilson coefficient $\xi_{T,n}$ are the same as for the product $m_q C_7$, where $C_7$ is the Wilson coefficient of the operator $Q_7\sim m_b(\bar{s}_L\sigma^{\mu\nu}b_R)F_{\mu\nu}$ in the $b\to s\gamma$ effective Lagrangian. Combining the matching coefficients of the quark masses given above~\cite{Chetyrkin:1997un} and the two-loop matching coefficient of $C_7$ given in Ref.~\cite{Misiak:2010sk}, we find
\bea
\xi_{T,1}=0 \,, \quad \xi_{T,2} = -\frac{97}{1296} \,.\label{eq:thrsld_tensor}
\eea

\section{RG running and SMEFT matching}
\label{sec:running}

Next, we solve the coupled differential RGE equations presented in the previous section, i.e. working at three-loop in QCD and one-loop in QED/EW. We work at the same order for the corresponding couplings constants $\alpha_s$, $\alpha_{em}$ and $g_{L,Y}$~\cite{Pich:1998xt}. Finite threshold corrections are also taken into account, both for the couplings and the Wilson coefficients.

\textbf{RG running below the weak scale.-} 
For the decays of light quarks, the low-energy Wilson coefficients $\epsilon_\Gamma$ are typically extracted from the experiment at $\mu=2$ GeV, simply because that is the renormalization scale chosen usually to extract the necessary lattice form factors. The subsequent running  to the weak scale is given by\footnote{For phenomenological applications it is convenient to use the (pseudo)scalar coefficients $\epsilon_{S/P}=\eSL\pm\eSR$ rather than their chiral counterparts. Their anomalous dimension matrix can be obtained by a trivial transformation of the matrices eq.~(\ref{eq:anomalous}).}
\bea
\left(
\begin{array}{c}
\eL\\
\eR\\
\eS\\
\eP \\
\eT \\
\end{array}
\right)_{\!\!\!\mbox{($\mu=2$~GeV)}}
\!\!\!\!\!= 
\left(
\begin{array}{ccccc}
 1 & 0 &0&0&0\\
 0 & 1.0046&0&0&0\\
 0 & 0 & 1.72 & 2.46\times 10^{-6} & -0.0242 \\
 0 & 0 & 2.46\times 10^{-6} & 1.72 & -0.0242 \\
 0 & 0 & -2.17 \times 10^{-4} & -2.17 \times 10^{-4}& 0.825 \\
 \end{array}
\right)
\!\!\left(
\begin{array}{c}
\eL\\
\eR\\
\eS\\
\eP \\
\eT \\
\end{array}
\right)_{\!\!\!\mbox{($\mu=Z$)}}
\label{eq:RGEepsilon}
\eea
where we took into account the bottom quark threshold. In order to illustrate the numerical importance of the QCD loops, let us focus on the $(3,3)$ element. Working at $n=\{0,1,2,3\}$ loops we find $\{1.00,1.51,1.69,1.72\}$. The 2-loop correction is clearly non-negligible, whereas the 3-loop one is comparable to the error introduced by $\alpha_s(m_Z)$~\cite{Olive:2016xmw}.

In the decay of bottom quarks, the reference scale is instead the $b$ quark mass $m_b$. Therefore the corresponding running to the weak scale is a bit smaller and it does not involve crossing any threshold. We find:
\bea
\left(
\begin{array}{c}
\eL\\
\eR\\
\eS\\
\eP \\
\eT \\
\end{array}
\right)_{\!\!\!\mbox{($\mu=m_b$)}}
\!\!\!\!\!=
\left(
\begin{array}{ccccc}
1 & 0 & 0 & 0 & 0\\
0 & 1.0038 &0&0&0\\
0 & 0 & 1.46 & 1.45\times 10^{-6} & -0.0177 \\
0 & 0 & 1.45\times 10^{-6}& 1.46 & -0.0177 \\
0 & 0 & -1.72 \times 10^{-4} &  -1.72 \times 10^{-4} & 0.878 \\
\end{array}
\right)
\!\!\left(
\begin{array}{c}
\eL\\
\eR\\
\eS\\
\eP \\
\eT \\
\end{array}
\right)_{\!\!\!\mbox{($\mu=Z$)}}\!\!\!.
\label{eq:RGEepsilon_b}
\eea

\textbf{Matching equations at the weak scale.-} 
Once the coefficients $\epsilon_\Gamma$ are evolved to the weak scale they can be matched to the proper EFT or model relevant at that scale. Here we matched them to SMEFT at order $\Lambda^{-2}$ and at tree level. This matching was first given in Ref.~\cite{Cirigliano:2009wk} using a specific operator basis, which was a slightly modified version of the Buchmuller-Wyler basis~\cite{Buchmuller:1985jz} with all relevant redundancies taken care of. Here we show the same equations, but this time using the popular Warsaw basis that was developed afterwards~\cite{Grzadkowski:2010es}
\bea
\frac{\delta G_F}{G_F}
&=& [w_{\varphi l}^{(3)}]_{11} + [w_{\varphi l}^{(3)}]_{22} - \frac{1}{2} [w_{ll}]_{1221},
\\
V_{ij} \cdot \eL^{ij\ell}
&=& V_{ij}  \,  [w_{\varphi l}^{(3)}]_{\ell\ell}   +  [w_{\varphi q}^{(3)}\,V]_{ij} -  [w_{l q}^{(3)}\, V]_{\ell\ell ij}, \nonumber\\
V_{ij} \cdot  \eR^{ij\ell}
&=& \frac{1}{2}[w_{\varphi ud}]_{ij}, 
\\
V_{ij} \cdot \epsilon_{S/P}^{ij\ell}
&=& - \frac{1}{2}[w_{\ell equ}^\dagger \,V \pm w_{ledq}^\dagger]_{\ell\ell ij}, 
\\
V_{ij} \cdot  \eT^{ij\ell}
&=& - \frac{1}{2}[(w^{(3)}_{\ell equ})^\dagger \,V ]_{\ell\ell ij}.
 \label{eq:matchingeqs}
\eea
Operators containing flavor indices are defined in the flavor basis where the up-quark Yukawa matrices are diagonal. Here repeated indices $i,j,\ell$ are not summed over; and transposition and matricial notation only affect quark indexes.

\textbf{RG running above the weak scale.-} 
In order to match to the underlying NP model, or to make contact with searches performed at high-energy colliders such as the LHC, it is necessary to perform an additional running from the weak scale to higher scales. Focusing once again on the chirality-breaking operators  we find:
\bea
\left(
\begin{array}{c}
w_{ledq} \\
w_{\ell equ}\\
w^{(3)}_{\ell equ} \\
\end{array}
\right)_{\!\!\!\mbox{($\mu=m_Z$)}}
= 
\left(
\begin{array}{ccc}
1.19 & 0. & 0. \\
 0. & 1.20 & -0.185 \\
 0. & -0.00381 & 0.959 \\
\end{array}
\right) 
\left(
\begin{array}{c}
w_{ledq}\\
w_{\ell equ} \\
w^{(3)}_{\ell equ} \\
\end{array}
\right)_{\!\!\!\mbox{($\mu=1$ TeV)}},
\label{eq:RGEsmeft}
\eea
where we took into account the top quark threshold.

\section{Phenomenology}
\label{sec:phenomenology}
\subsection{Light quark decays}
\label{sec:lightquarks}

In Ref.~\cite{Gonzalez-Alonso:2016etj} the $d(s)\to u \ell \bar{\nu}_\ell$ transitions, such as nuclear, baryon and meson decays, were studied within the SMEFT and obtained bounds for 14 combinations of effective low-energy couplings between light quarks and leptons ($\epsilon_\Gamma^{du\ell}$) at the hadronic scale $\mu=2$ GeV. Using the results obtained in the previous section we can evolve them to the weak scale where they can be matched to the SMEFT coefficients. The resulting bounds and correlations are given in~\aref{appA}. 

In order to illustrate the importance of the QED mixing, let us investigate a NP scenario where only the tensor interaction $\eT$ is generated at the weak scale.\footnote{Explicit models that produce tensor operators are for example those including tree-level leptoquark-mediated interactions, although in this case (pseudo)scalar operators are also generated~\cite{Davidson:1993qk,Sakaki:2013bfa,Alonso:2015sja,Freytsis:2015qca}.}  
\fref{barchart} shows the bounds on such an interaction obtained from the phenomenological analysis of Ref.~\cite{Gonzalez-Alonso:2016etj} neglecting and including the electromagnetic mixing. 
As we can see, the inclusion of the mixing increases the bound by several orders of magnitude in some cases. Needless to say, the effect is even larger if one chooses a higher scale (instead of the weak scale) to assume the dominance of the tensor interaction, not only because of the larger running but also because the EW mixing is larger than the QED one. 
\begin{figure}[h!]
\centerline{\includegraphics[scale=0.80]{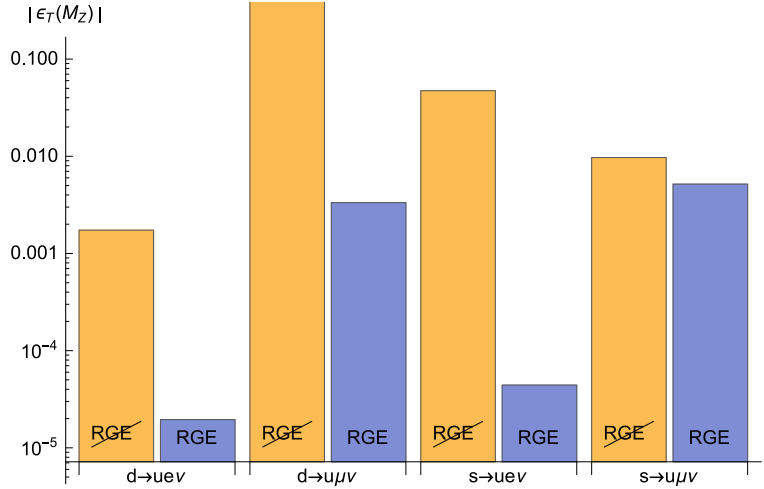}}
\caption{90\% C.L. bounds on the tensor operator $\eT(m_Z)$ for different flavor structures. The yellow (blue) bars show the results neglecting (including) the QED mixing. The QCD running is included in both cases. Only one operator at a time is present. The $\eT^{ud\mu}  (m_Z)$ coefficient is unbounded if the mixing is neglected, which is represented by an unlimited vertical bar.
\label{fig:barchart}}
\end{figure}

The reason of this is well known: the bound obtained at the low-energy scale on the pseudoscalar coupling $\eP$ is usually much stronger than the bounds over the scalar or tensor couplings. This is so because the pseudoscalar coupling contributes at tree level to the leptonic meson decay without the helicity suppression that affects the SM contribution, namely~\cite{Gonzalez-Alonso:2016etj}
\bea
\frac{\Gamma (P\to \ell\nu_\ell)}{\Gamma_{\rm SM}}
&=& 
\left| 1+\eL^{ij\ell}-\eR^{ij\ell} -\frac{m_{P^\pm}^2}{m_\ell\left(m_j+m_i\right)}\eP^{ij\ell}\right|^2
\eea
where $(i,j)=(u,d),(u,s)$ for $P=\pi,\,K$ respectively. The NP sensitivity of these channels is further enhanced by very precise measurements and theoretical predictions, especially in the lepton-universality ratios $R_P=\Gamma(P_{e2(\gamma)})/\Gamma(P_{\mu2(\gamma)})$~\cite{Olive:2016xmw,Aguilar-Arevalo:2015cdf,Cirigliano:2007xi}. 
This makes possible to set limits at the $10^{-7}$ level on the electronic couplings $\eP^{ude}$ and $\eP^{use}$~\cite{Gonzalez-Alonso:2016etj}, corresponding to a NP effective scale of ${\cal O}(500)$ TeV. Such enormous sensitivity largely compensates for the 1-loop suppression that affects the tensor coefficient $\epsilon_T$ and makes possible to set the strong bounds shown in \fref{barchart}. Furthermore, as we will see below, this data can also be used to obtain indirect bounds on operators probing different flavor structures such as those in $b$-quark decays, where the high sensitivity to the pseudoscalar Wilson coefficient compensates now for CKM-suppression factors.

\subsection{Heavy-quark decays}

New physics scenarios involving (pseudo)scalar and tensor operators to explain the $B$-decays anomalies in the $b\to (c,\,u) \ell\nu$ transitions are phenomenologically viable and have been explored extensively in the literature~\cite{Biancofiore:2013ki,Sakaki:2014sea,Duraisamy:2014sna,Freytsis:2015qca,Becirevic:2016hea,Alonso:2016gym,Colangelo:2016ymy,Li:2016vvp,Bardhan:2016uhr,Bhattacharya:2016zcw,Ivanov:2017mrj,Chen:2017hir}. \footnote{The tensions in $|V_{cb}|$ could be solved within the SM by reanalyzing the parametrizations of the form factors in the exclusive $B\to D^*$ decay mode, as has been shown recently in refs.~\cite{Bigi:2017njr,Grinstein:2017nlq}. Note that our analysis in the EFT of NP is completely general; we will adopt in the following, and for the sake of the argument, the results in ref.~\cite{Colangelo:2016ymy} as a benchmark for a tensor NP scenario contributing to the $B\to D^{(*)}\ell\nu$ decays.} In this section we show that the effect of the RGE mixing between operators, which was not taken into account in these studies, can be important. 

Let us first focus on the analysis of the $b\to c\ell\nu$ transitions. The $R_{D^{(*)}}$ and $|V_{cb}|$ anomalies can be accommodated with the contributions
\begin{align}
\eT^{c b \tau}(m_b)\simeq0.38~~\text{\cite{Alonso:2016oyd},}~~~~~~~~~~~~~~~~~~~\eT^{c b (e,~\mu)}(m_b)\simeq0.2~~\text{\cite{Colangelo:2016ymy},}
\end{align}
respectively, which in the SMEFT stem from the tensor operator ${\cal O}_{\ell e q u}^{(3)}$ in Tab.~\ref{tab:warsaw}. For  simplicity, we assume that only the entry in flavor space connecting the second and third quark families of this operator is nonzero. Taking into account the running 
 we find that these contributions to the tensor operators at $\mu=m_b$ can be generated by
\begin{align}
 [w_{\ell e qu}^{(3)}]_{\tau\tau 32}(\Lambda)=-0.038,~~~~~~~~~~~~~~~~~~~[w_{\ell e qu}^{(3)}]_{\ell \ell 32}(\Lambda)=-0.02,\label{eq:tensNPsc}
\end{align}
at the NP scale $\Lambda=1$ TeV. However, even if only that operator was generated in the matching to the NP model, it will induce by mixing under EW interactions (\emph{cf.} \eref{RGEsmeft}) the following contribution to the scalar operators at the scale of the $Z$-mass 
\begin{align}
 [w_{\ell e qu}]_{\tau \tau 32}(m_Z)=0.0071,~~~~~~~~~~~~~~~~~~~[w_{\ell e qu}]_{\ell \ell 32}(m_Z)= 0.0037.\label{eq:tensZmc}
\end{align} 
Matching the SMEFT and the low-energy EFT at that scale using Eqs.~(\ref{eq:matchingeqs}) and running down to $\mu=m_b$ using \eref{RGEepsilon_b}, we find the contributions
\begin{align}
\eS^{c b \tau}(m_b)=\eP^{c b \tau}(m_b)=-0.129,~~~~~~~~~~~~~~~~~~ \eS^{c b \ell}(m_b)=\eP^{c b\ell}(m_b)=-0.068.\label{eq:Scalarsmbsc}
\end{align}

These contributions to the scalar operators at low energies are significant. Let us first look at their effects on the lepton-universality ratios $R_D$ and $R_{D^*}$, which are summarized in the following table:

\begin{center}
\begin{tabular}{|c|cccc|}
\hline
&expt.&SM& Mixing neglected & Mixing included\\
\hline
$R_D$&$0.397(40)(28)$&$0.310$&$0.458$&$0.395$\\
$R_{D^*}$&$0.316(16)(10)$&$0.252$&$0.312$&$0.316$\\
\hline
\end{tabular}
\end{center}

The first two columns show the current experimental averages and SM predictions~\cite{Alonso:2016gym}. The third column shows the $R_{D^{(*)}}$ values generated by the tensor contribution in the left-hand side of \eref{tensNPsc}, taking into account the QCD running but neglecting the QED and EW mixing,
\bea
[w_{\ell e qu}^{(3)}]_{\tau\tau 32} (1~{\rm TeV})=-0.038 ~~~~\to~~~~ (\eS^{c b \tau},\eP^{c b \tau},\eT^{c b \tau})(m_b)=(0,0,0.380)~.
\eea
Finally, the fourth column shows the values generated by the same initial operator at the high energy scale (1 TeV), but this time including also the effect of the (pseudo)scalar interactions produced through the QED and EW mixing, \emph{cf.}~\eref{Scalarsmbsc},
\bea
[w_{\ell e qu}^{(3)}]_{\tau\tau 32} (1~{\rm TeV})=-0.038 ~~~~\to~~~~ (\eS^{c b \tau},\eP^{c b \tau},\eT^{c b \tau})(m_b)=(-0.129,-0.129,0.380)~.
\eea  
The $R_{D^{(*)}}$ values including non-standard effects in the above-given table are obtained following Ref.~\cite{Alonso:2016gym}. As we can see, the impact of the RGE mixing in the $B\to D\tau\nu$ predictions is important due to the well known sensitivity of this channel to the scalar operator, contrary to the case of the $B\to D^*\tau\nu$ channel, which has little sensitivity to the pseudoscalar contribution. Interestingly enough, we find that the inclusion of mixing effects improves significantly the agreement of the tensor scenario with data. Models for which these mixing effects are relevant include leptoquarks, such as those in Refs.~\cite{Sakaki:2014sea,Freytsis:2015qca,Li:2016vvp,Chen:2017hir}. New physics producing scalar contributions at the TeV or electroweak scales, such as two-Higgs doublet models (see e.g. Ref.~\cite{Celis:2016azn}), experience mixing into the tensor operator. However, as it can be concluded from the corresponding entries in the matrices~(\ref{eq:RGEepsilon_b}) and (\ref{eq:RGEsmeft}), this has a very small impact in the phenomenology.  

On the other hand, the RGE-induced pseudoscalar operator also modifies the branching fraction of the decay $B_c\to\tau\nu$ with a contribution that does not suffer the helicity suppression, namely
\bea
BR(B_c\to\tau\nu) \approx\, 0.06~,
\eea
which is a factor 2.5 larger than the SM branching fraction, $BR(B_c\to\tau\nu)|^{\rm SM}=2.4\%$. We used PDG data for the various masses~\cite{Olive:2016xmw} and the $f_{B_c}$ lattice result from Ref.~\cite{Colquhoun:2015oha}. This is an important enhancement, although not as large as to enter in conflict with the bounds that can be obtained from the lifetime of the $B_c$ and the calculations of the corresponding inclusive decay width in QCD~\cite{Alonso:2016oyd}. 

Next we investigate the effects of the tensor contribution on the light-lepton operators, i.e. the right-hand side of \eref{tensNPsc}
\bea
[w_{\ell e qu}^{(3)}]_{\ell \ell 32}(1~{\rm TeV})=-0.02 ~~~~\to~~~~ (\eS^{c b \ell},\eP^{c b \ell},\eT^{c b \ell})(m_b)=(-0.068,-0.068,0.200)~.
\eea  
In contrast to the $B_c\to\tau\nu$ decay, and more in line with the pion and kaon decays above, the contributions of the RGE-induced pseudoscalar operator to $B_c\to\ell\nu$ can be much more amplified due to the chiral enhancement. Namely
\begin{align}
BR(B_c\to\mu\nu)\approx 3.3\times10^{-3},~~~~~~~~~~~~~~BR(B_c\to e\nu)\approx 2.3\times10^{-3},
\end{align}
which correspond to an enhancement of one order of magnitude for the muonic decays with respect to the SM ($BR(B_c\to\mu\nu)|^{\rm SM}=1.0\times10^{-4}$) and no less than six orders of magnitude for the electronic case ($BR(B_c\to e\nu)|^{\rm SM}=2.3~\times10^{-9}$). Although the branching fraction is smaller than in the $\tau$ channel, the experimental signal is cleaner in this case. 
All in all, the leptonic decay modes of the $B_c$ meson represent an interesting tool to probe the tensor scenarios, with predictions that could be at reach for the LHC in the near future.

\subsubsection{Indirect bounds from light-quark decays}

The leptonic decays of the light-quarks can probe SMEFT coefficients with flavor structures involving third quark generation even if their contribution is suppressed by CKM factors arising when transforming to the mass basis. In order to illustrate this let us now focus on possible contributions of chirality-flipping NP operators in the $b\to u \ell\nu$ transitions, which could be advocated to, e.g., solve the inclusive versus exclusive discrepancy on $|V_{ub}|$~\cite{Amhis:2014hma}. Let us take, for example, a NP model whose main contribution is the (real) Wilson coefficient $[w_{\ell equ}]_{ee31}$, in our SMEFT flavor basis. Such an operator will generate, according to the matching equations~(\ref{eq:matchingeqs}), a leading contribution to the $b\to u e\nu$ decays. Namely,
\begin{align}
\epsilon_{S/P}^{ube}(m_Z)=-\frac{V_{tb}}{2V_{ub}}[w_{\ell equ}]_{ee31}(m_Z).
\end{align}
However it also gives a contribution to $s\to u e\nu_e$ transitions via, 
\begin{align}
\epsilon_{S/P}^{use}(m_Z)=-\frac{V_{ts}}{2V_{us}}[w_{\ell equ}]_{ee31}(m_Z),
\end{align}
which, despite the strong CKM suppression, can lead to very strong bounds from kaon decay data, as discussed in \sref{lightquarks}. For instance, the process $B\to e\nu$ has an experimental limit of $\leq10^{-6}$ at $90\%$ confidence level in the branching fraction~\cite{Olive:2016xmw}. Taking into account the running, this translates into the following bound for the underlying SMEFT Wilson coefficient
\begin{align}
 \left|[w_{\ell equ}]_{ee31}(m_Z)\right|\leq 1.3\times10^{-4}~,\label{eq:buenu_direct}
\end{align}
where we use the $N_f=2+1+1$ lattice result $f_B=0.186(4)$ GeV~\cite{Dowdall:2013tga}.
On the other hand, applying the bounds discussed above from kaon electronic decays~\cite{Gonzalez-Alonso:2016etj} we obtain a much stronger bound
\begin{align}
[w_{\ell equ}]_{ee31}(m_Z)=(2.9\pm 4.0)\times10^{-6},	\label{eq:buenu_indirect}
\end{align}
at $90\%$ C.L. Similar conclusions can be derived for tensor operators through the mixing discussed in the previous sections. 

Let us stress, to finish, that the purpose of this simple exercise, with only one coefficient ($[w_{\ell equ}]_{ee31}$) at the electroweak scale, is illustrative. The specific numbers will certainly change if more operators or flavor-space entries are included, which will be probably the case for any realistic NP model or in scenarios advocating specific flavor symmetries. Nonetheless, the main message remains: NP models (or SMEFT benchmark scenarios) addressing the so-called $B$ anomalies could be potentially constrained by light quark decays.

\section{Conclusions}
\label{sec:conclusions}

We studied the renormalization group evolution of (pseudo)scalar and tensor new physics contributions to the (semi)leptonic charged-current meson decays. We calculated the mixing and evolution under electroweak and electromagnetic interactions, including also the three-loop running in QCD and taking into account the crossing of quark thresholds. We provided numerical formulas, \emph{cf.} Eqs.~(\ref{eq:RGEepsilon}), (\ref{eq:RGEepsilon_b}) and (\ref{eq:RGEsmeft}), which connect the values of the corresponding Wilson coefficients at the high-energy scales, relevant for collider physics and model-building, with their values at the low energies where they are phenomenologically accessible in meson decays. These results can be be trivially implemented in public codes~\cite{Gonzalez-Alonso:2016etj,flavio,Celis:2017hod,Workgroup:2017myk} and in future analysis of these transitions. As an example, we used our results to evolve the bounds obtained in the recent global fit of Ref.~\cite{Gonzalez-Alonso:2016etj} using nuclear, pion and kaon decay data at the low-energy scale to the electroweak scale.

From the phenomenological point of view, the most important effect is related to the large mixing of the tensor operator into the scalar ones, both through QED and electroweak one-loop effects. We first illustrated the importance of this mixing for light quark decays, discussing the stringent bounds that can be derived on tensor operators by exploiting the mixing and the fact that the induced (pseudo)scalar operators lift the helicity suppression that purely-leptonic decays of pions and kaons receive in the SM. 

We also showed how this helicity suppression can also be exploited in tensor scenarios that address the exclusive $vs.$ inclusive determinations of $|V_{cb}|$ and $|V_{ub}|$. Indeed, the RGE-induced pseudoscalar operators would produce enhancements with respect of the SM that in the case of the $B_c\to e\nu$ can reach several orders of magnitude. 

Furthermore, we showed that the RGE mixing has an important impact on the tensor explanations of the $R_{D^{(*)}}$ anomalies. The sizable scalar contributions generated radiatively at low energies modify importantly the phenomenology and, interestingly enough, improves the agreement of this scenario with data. Finally, we discussed the indirect bounds that can also be derived on operators connecting to third-quark generation using the very stringent bounds stemming again from the leptonic pion and kaon decays. 

We conclude by noting that other finite radiative corrections are expected in the maching of specific NP models to the SMEFT, the matching between the SMEFT and the low-energy EFT, and from the radiative corrections to the physical processes. In comparison, the pieces computed in this work receive logarithmic enhancements which do not represent a large amplification if the NP scale is not very high. Our work shows that radiative corrections can be important for the low-energy phenomenology of the chirally-flipping operators and should be taken into account in new physics studies of the (semi)leptonic meson decays.

\textit{Note added.-} As this work was being completed, Ref.~\cite{Aebischer:2017gaw} appeared, including the 1-loop QED anomalous dimension that agrees with our result of \eref{gammaqed}. Our work in progress was presented by one of us (M.G.-A.) at the Portoroz 2017 workshop on April 20, 2017.
%
\section*{Acknowledgements}

We thank R.~Alonso, V.~Mateu, M.~Misiak and A.~Pich for useful discussions, and A.~Falkowski for his valuable collaboration in the early stage of this work. M.G.-A. is grateful to the LABEX Lyon Institute of Origins (ANR-10-LABX-0066) of the Universit\'e de Lyon for its financial support within the program ANR-11-IDEX-0007 of the French government.

\appendix

\section{Results of the fit to light quarks at $\mu=m_Z$}
\label{app:appA}

Here we present the result of the fit to nuclear, hyperon, pion and kaon decay data of Ref.~\cite{Gonzalez-Alonso:2016etj} evolved from $\mu=2$ GeV to $\mu=m_Z$ using eq.~(\ref{eq:RGEepsilon}). We refer the reader to this reference for details on notation and definitions. We find
\bea
\left(
\begin{array}{c}
 \Delta_{\rm CKM} \\
 \eL^{us\mu}-\eL^{use} \\
 \eL^{ud\mu}-\eL^{ude}- 41\, \eP^{ud\mu} + 0.58\, \eT^{ud\mu} \\
 \eR^{ude}\\
 \eR^{use}\\
  \eS^{ude} \\
 \eP^{ude} \\
 \eT^{ude} \\
 \eS^{use} \\
 \eP^{use} \\
 \eT^{use} \\
 \eS^{us\mu} \\
 \eP^{us\mu} \\
 \eT^{us\mu} \\
\end{array}
\right)_{\!\!\!\!\mu=m_Z}
=
\left(
\begin{array}{c}
 -1.2\pm 8.4 \\
 1.0\pm 2.5 \\
 -1.9\pm 3.8 \\
 -1.3\pm 1.7 \\
 0.1\pm 5.0 \\
 8.0\pm 7.5 \\
 0.4\pm 1.4 \\
 1.2\pm 9.7 \\
 -0.8\pm 1.9 \\
 1.5\pm 3.1 \\
 1.1\pm 2.2 \\
 -2.2\pm 3.0 \\
 -0.4\pm 2.5 \\
 0.6\pm 6.3 \\
\end{array}
\right)\times 10^{\wedge}\left(
\begin{array}{c}
 -4 \\
 -3 \\
 -2 \\
 -2 \\
 -2 \\
 -4 \\
 -5 \\
 -4 \\
 -3 \\
 -4 \\
 -2 \\
 -4 \\
 -3 \\
 -3 \\
\end{array}
\right)
\eea
where 
\bea 
\label{eq:deltackm1}
\Delta_{\rm CKM}  & =  & 
2|V_{ud}|^2(\eL^{ude}+\eR^{ude})+2|V_{us}|^2(\eL^{use}+\eR^{use})-2\frac{\delta G_F}{G_F}~.
\eea
The correlation matrix given by
\bea
\rho = \scriptsize{
\left(
\begin{array}{cccccccccccccc}
 1. & -0.07 & 0.01 & 0. & 0. & 0.73 & 0. & 0. & 0. & 0. & 0. & -0.11 & 0.02 & 0. \\
 - & 1. & 0. & 0. & 0. & 0. & 0. & 0. & 0. & 0. & 0. & 0.14 & 0.04 & 0.46 \\
 - & - & 1. & -0.87 & 0. & 0.01 & 0.32 & 0. & 0. & 0. & 0. & 0.04 & 0.09 & 0. \\
 - & - & - & 1. & 0. & 0. & -0.27 & 0. & 0. & 0. & 0. & 0. & 0. & 0. \\
 - & - & - & - & 1. & 0. & 0. & 0. & 0. & -0.04 & 0. & 0. & -0.98 & 0. \\
 - & - & - & - & - & 1. & 0.02 & 0.02 & 0. & 0. & 0. & 0. & 0. & 0. \\
 - & - & - & - & - & - & 1. & 0.95 & 0. & 0. & 0. & 0.01 & 0.03 & 0. \\
 - & - & - & - & - & - & - & 1. & 0. & 0. & 0. & 0. & 0. & 0. \\
 - & - & - & - & - & - & - & - & 1. & 0.16 & 0.16 & 0. & 0. & 0. \\
 - & - & - & - & - & - & - & - & - & 1. & 0.9992 & 0. & 0.04 & 0. \\
 - & - & - & - & - & - & - & - & - & - & 1. & 0. & 0. & 0. \\
 - & - & - & - & - & - & - & - & - & - & - & 1. & 0. & 0.3 \\
 - & - & - & - & - & - & - & - & - & - & - & - & 1. & 0.05 \\
 - & - & - & - & - & - & - & - & - & - & - & - & - & 1. \\
\end{array}
\right)}
\eea

\bibliographystyle{JHEP}
\bibliography{rgpaper.bib}
\end{document}